\documentclass[twocolumn]{openjournal_v3}
\usepackage{graphicx} 

\usepackage[colorlinks,linkcolor=blue,citecolor=blue,urlcolor=blue ]{hyperref}
\usepackage[utf8]{inputenc}
\usepackage{float}
\usepackage{xcolor}
\usepackage{ulem}
\usepackage[T1]{fontenc}
\usepackage{amsmath}
\usepackage{xparse}
\usepackage{physics}

\newcommand{\msolaryr} {$\rm{M_{\odot} \ yr^{-1}}$ }
\newcommand{\renaissance}{\texttt{Renaissance}}

\newcommand{\msolar}{\rm{M}_\odot}

\begin{document}

\title{Beyond No Tension: JWST $z > 10$ Galaxies Push Simulations to the Limit\vspace{-1.5cm}}

\author{Joe McCaffrey$^1$}
\author{Samantha Hardin$^2$}
\author{John H. Wise$^2$}
\author{John A. Regan$^1$}
\affiliation{$^1$Centre for Astrophysics \& Space Science Maynooth, Maynooth University, Maynooth, Co. Kildare, Ireland\\
$^2$Center for Relativistic Astrophysics, Georgia Institute of Technology, 837 State Street, Atlanta, GA 30332, USA}
\thanks{joe.mccaffrey@mu.ie}

\begin{abstract}
    \noindent JWST has identified some of the Universe’s earliest galaxies, repeatedly pushing the frontier to ever higher redshifts and stellar masses. The presence of such extreme galaxies at such early times, with large stellar populations and high star-formation rates, naturally results in a tension between observation and theory. This tension between numerical models and observations can be either due to our underlying cosmological models or due to a gap in our understanding of early Universe astrophysics.

    In a prelude to this letter, we showed how the \renaissance{} simulations, which focused on high redshift galaxy formation were able to reconstruct similar stellar masses to the earliest and highest mass galaxies that had been discovered by JWST at the time of its publication \citep{mccaffreyNoTensionJWST2023c}. Since then many more galaxies have been discovered by JWST, in particular the ``Mirage-or-Miracle'' (MoM) survey broke the record recently with the highest redshift galaxy MoM-z14, which has a spectroscopically confirmed redshift of $z\sim14.44$ followed closely by GS-z14 with a spectroscopically confirmed redshift of $\rm{z} \sim 14.18$. We investigate in this letter whether these newly discovered galaxies are in conflict with the \renaissance{} simulations and thus whether they are causing tension with our established models of cosmology and/or high-redshift astrophysics. We discover that MoM-z14's high mass at early redshift can be explained by the Renaissance simulation suite, whereas the extremely high stellar mass of GS-z14 remains an outlier when compared to previous measurements of high-redshift galaxies detected by JWST and our numerical models (even after accounting for cosmic variance).
    
\end{abstract}

\maketitle

\section{Introduction}

\begin{figure*}[!htp]
    \centering
    \includegraphics[width=0.9\textwidth]{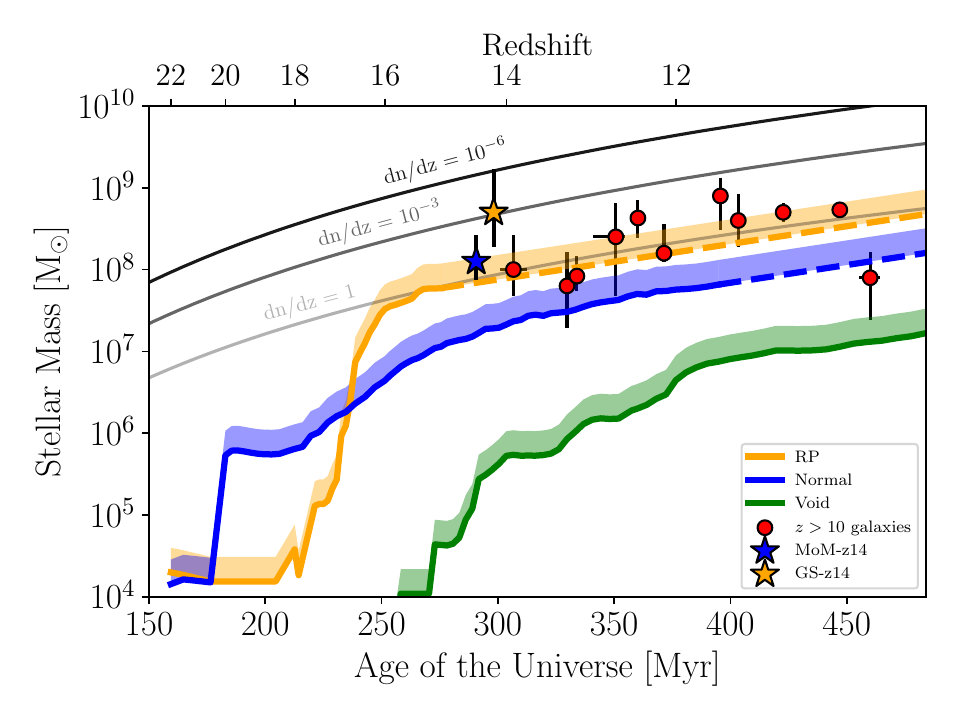}
    \caption{The growth evolution of the most massive galaxies from each region in \renaissance{} compared with observations from JWST. The dashed lines indicate where the stellar mass is extrapolated after the end time of each region in \renaissance{}, using a specific star formation rate of $10^{-8}$. The shaded regions denote the upper limit to these galaxies if we take into account an error from cosmic variance of $+100\%$. The greyscale lines denoted by $dn/dz$ indicate the probability of discovering a galaxy of a certain stellar mass and at a certain redshift in the FoV of the NIRCam instrument on JWST. The calculation of these lines can be found in Appendix A of \cite{mccaffreyNoTensionJWST2023c}.}
    \label{fig:stellar-mass}
\end{figure*}

\begin{figure*}[!htpb]
    \centering
    \includegraphics[width=0.9\textwidth]{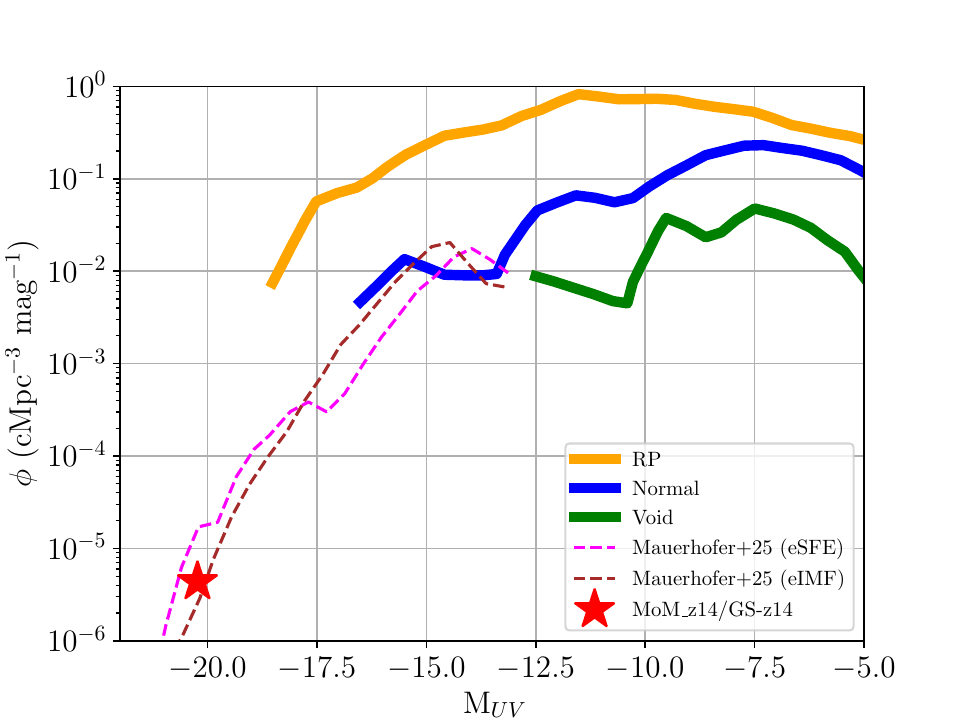}
    \caption{Luminosity function of the galaxies modelled in \renaissance{}, along with the luminosity function derived from the parameters constrained from GS-z14 and MoM-z14 \citep{naiduCosmicMiracleRemarkably2025}. We have also included estimates of the luminosity function derived from models which assume a top-heavy IMF in the early Universe (brown line) and a model which assumes that star formation efficiencies increase with increasing redshift \citep{mauerhoferSynergisingSemianalyticalModels2025}. It is noted that the luminosity function derived by \cite{naiduCosmicMiracleRemarkably2025} assumes that these galaxies are the only $-21 < M_{\rm UV} < -20$ galaxies to exist at $z\sim14-15$ in the UDS, GOODS-S and COSMOS fields, imaged by the PRIMER and JOF surveys.}
    \label{fig:lf}
\end{figure*}

\noindent It has been three years since JWST started providing fresh information on previously hidden parts of the Universe. From exoplanets to the extragalactic population, we have gained unprecedented access into astrophysics that previously remained theoretical. In this paper specifically, we are focused on understanding the formation of the very first galaxies and how accurate well established cosmological models are at predicting the number densities and formation rates that these galaxies would exhibit. Not long after the commissioning of JWST that probes the high redshift Universe, questions started to arise into what observations were beginning to tell us about early galaxy formation. Surveys like JADES \citep{bunkerSpectroscopyJWSTAdvanced2019, JADES-2, JADES-3} and CEERS \citep{CEERS-2,CEERS, CEERS-3} unveiled a population of galaxies that appeared to even violate accepted formation models.  They displayed high stellar masses and high star formation rates for this epoch of the Universe, reaching the order of $10^8 - 10^9 \ \msolar$ in stellar mass only a few hundred million years after the Big Bang \citep{robertson, curtis-lakeSpectroscopicConfirmationFour2023a}.  These galaxies presented a problem as theoretical models predict that if these galaxies do in fact exist, they would have to possess star formation efficiencies close to 100\%, given the accepted baryonic matter made available by $\Lambda$CDM -- our most accepted cosmological model at the time of writing this letter \citep{boylan-kolchinStressTestingLCDM2023a}. \\
\indent Resolutions to this tension include the possibility that the stellar masses from early galaxies are being overestimated, due to possible contribution of light from massive black holes in these high-z galaxies \citep[e.g.][]{chworowskyEvidenceShallowEvolution2024}. Other research lines find that for high mass dark matter halos, that star formation becomes more efficient with redshift, allowing for the existence of high stellar masses in these halos without violating $\Lambda$CDM \citep[e.g.][]{shuntov-sfe}. 

One way to see what kind of galaxy population $\Lambda$CDM would present us with is the use of cosmological simulations. With simulations we are able to construct a Universe using a priori assumptions based on the most basic physics with initial conditions provided by measurement of the Cosmic Microwave Background and also sophisticated state-of-the-art modelling of physical mechanisms. Due to current computational power, simulations are often limited in terms of the redshift regimes that are being examined along with the mass and spatial resolutions. To quantify this aspect of numerical work \cite{keller2022} utilised a number of cosmological simulations to try and see if they could reconstruct the high-z galaxies that were observed being with JWST. While these simulations were able to produce galaxies on the same mass scale  as the observations, they were not designed to focus on the high-z galaxies, and so the build up of these galaxies are not accurately captured by the large scale cosmological simulations such as EAGLE \citep{joop-eagle, mcalpineEagleSimulationsGalaxy2016}, Illustris \citep{vogelsbergerPropertiesGalaxiesReproduced2014}, TNG100 \citep{naimanFirstResultsIllustrisTNG2018}, RomulusC  \citep{tremmelIntroducingRomuluscCosmological2019}, Obelisk \citep{obelisk} and Simba \citep{daveSimbaCosmologicalSimulations2019}. In \cite{mccaffreyNoTensionJWST2023c} (hereafter M23), we used the \renaissance{} simulations \citep{osheaPROBINGULTRAVIOLETLUMINOSITY2015a, smithGrowthBlackHoles2018a, wiseFormationMassiveBlack2019}, a suite of high-z, zoom-in simulations that were specifically designed for modelling early Universe galaxy formation. With \renaissance{}, we were not only able to reproduce the high masses and star formation rates of the JADES/CEERS galaxies, but we were also able to show the build up of these galaxies from small mass scales due to the superior mass and spatial resolutions of \renaissance{}. In M23 we concluded that the existence of these high-z observations were not in tension with $\Lambda$CDM and that more evidence would be needed to provide a strong argument that theoretical models needed revisions.

Fast forward to 2025 and JWST has continued to break the record of spectroscopically confirmed high-z galaxies. Most recently, the Miracle-or-Mirage (MoM) survey has discovered a galaxy, MoM-z14 , in the COSMOS Legacy field which has a spectroscopically confirmed redshift of $z\sim14.44$ and a measured stellar mass of $10^{8.1} \ \msolar$ \citep{naiduCosmicMiracleRemarkably2025}. The galaxy also exhibits a luminous UV signature suggesting high levels of star formation for this epoch of its lifetime. Along with MoM-z14, the JADES survey has unveiled a galaxy at $z\sim14.18$, GS-z14, which has a mean spectroscopically derived stellar mass of $10^{8.6}\ \msolar$ \citep{GS-z14}, even higher than its high redshift counterpart MoM-z14. These extremely high-z and rapidly forming galaxies force us to revisit the work of M23 and deduce whether or not these galaxies are in direct tension with our cosmological simulations, and if so, do we need to revisit models of star formation/feedback or do we need to adjust how we apply models to observations of high-z environments. Complimentary research to ours has been undertaken by \cite{kohandelAmaryllisDigitalTwin2025a} to explain the existence of GS-z14. \cite{kohandelAmaryllisDigitalTwin2025a} used the SERRA (citation) zoom-in simulations to track the evolution of an analogous simulated galaxy to $z\sim 7$. They were able to show that their ``galaxy twin'' has characteristics including stellar mass, SFR and far infrared (FIR)\& UV luminosity lines comparable to GS-z14. They explain the existence of these luminous FIR lines as being transient events caused by merger-driven star formation episodes.

In this letter, we are updating the results of M23 and comparing the galaxies from \renaissance{} to the highest redshift observations from JWST, along with other $z>10$ galaxies to provide context on the steadily forming census of early galaxies. We also examine the measured UV luminosity functions from these $z\sim 14$ galaxies and see how it compares to the derived luminosity functions from the \renaissance{} simulations.

\section{Methodology}
\noindent We will now briefly outline the methods we applied to compare the JWST observations with the results from \renaissance{} simulations. First we will describe the main attributes of the \renaissance{} simulations which are relevant to this work, more information on \renaissance{} can be found in \citet{xuPopulationIIIStars2013b}, \citet{osheaPROBINGULTRAVIOLETLUMINOSITY2015a}, \citet{smithGrowthBlackHoles2018a}, and \citet{wiseFormationMassiveBlack2019}. \renaissance{} is a suite of N-body hydrodynamic simulations created using the adaptive mesh-refinement code Enzo \citep{Enzo_2014, Enzo_2019}. \renaissance{} consists of three zoom-in regions, defined by their level of overdensity: the Rarepeak (RP), Normal, and Void region. The RP region consists of a box of 3.8×5.4×6.6 cMpc$^3$ run to $z = 15$, while the Normal and Void regions consists of boxes of dimensions 6.0 × 6.0 × 6.125 cMpc$^3$, respectively run to $z = 11.6$ and $z = 8$. The advantages of \renaissance{} include its high mass resolution of dark matter particles, reaching down to $2.4 \times 10^4\ \msolar$ and spatial resolution reaching down to $19$ cpc. \renaissance{} also contains prescriptions for PopIII and PopII star formation and feedback, allowing for accurate galaxy formation modelling at high redshifts. Star formation in \renaissance{} relies on the stochastic sampling of the first generation of stars, based on individual star forming clouds which can be resolved in the simulations. A star formation efficiency of 7\% is used to convert cold gas within the simulations into PopII stars, which is based on simulations of local star formations simulations by \citep{SFE_renaissance}, who calibrated feedback models against local dwarf galaxies. This was to ensure that the simulations avoided the ``over-cooling problem'' which led to overestimation of stellar mass and metallicity. However, we note that this 7\% is not representative of the galaxy as a whole and that it is possible for the galaxy to achieve SFE's greater than this value. In fact from \cite{Xu_2016} it can be seen that the majority of halos experience a mean stellar baryonic fraction of $10^{-2}$.

We will use the results from the simulations to compare with the JWST measurements of MoM-z14 and GS-z14 and a number of other $z > 10$ galaxies taken from \cite{curtis-lakeSpectroscopicConfirmationFour2023a, bunker2023jades, wangUNCOVERIlluminatingEarly2023, haroConfirmationRefutationVery2023, bakxDeepALMARedshift2023, carnianiSpectroscopicConfirmationTwo2024}. Due to the high redshifts of these galaxies, we now have the chance to almost directly compare observations of high-z galaxies with like counterparts of galaxies found in the RP region of \renaissance{}, which ends at $z = 15$, only tens of millions of years of separation away from the observations. We will analyse the observations by studying them as individual cases, with the masses being compared to the most massive galaxies in \renaissance{}, but we will also try to compare these exceptional galaxies against the simulated global population of \renaissance{}, by connecting their calculated UV luminosity functions against the luminosity functions derived from the simulations, analysing the star formation models in the process.

\section{Results}

\noindent In Figure \ref{fig:stellar-mass} we plot the new stellar mass observations (MoM-z14 and GS-z14) against the values discussed in \cite{mccaffreyNoTensionJWST2023c}. On top of each solid line, which denotes the stellar mass of the most massive galaxy in each \renaissance{} realisation we include a region of error on the stellar mass that can be attributed to cosmic variance of +100\%, which we obtained from \cite{cosmic_variance}. We only include +100\% to provide an upper limit to the stellar mass of the galaxies, in order to show the upper limits that the most massive galaxy from each region can reach due to cosmic variance. The dotted lines following the RP and Normal solid lines extrapolate the stellar masses based on the star formation rates found for those galaxies at the ending redshifts. \\
\indent  MoM-z14 appears to follow the established trend of the previously discovered $z>10$ galaxies, with its stellar mass lying safely within the predicted mass that the most massive galaxy of the RP region. Only three of the discovered galaxies have mass ranges that intersect the most massive galaxy in the Normal region, despite these galaxies likely coming from a Normal-like environment, suggesting some revisions in stellar formation modelling at this epoch is required. In contrast, GS-z14 appears to be testing the limits of \renaissance{}, with its stellar mass lying above the accepted range of simulated masses at that redshift even for the RP region. This anomalous nature of GS-z14 initially suggests that \renaissance{} is underestimating the star formation in select high mass environments at high-redshift, which may be resolved in future through higher mass and spatial resolutions and updates to star formation and feedback prescriptions in such environments. \\
\indent We note however that this is one measurement that does not agree with simulations. This outlier, combined with the fact that JWST will only be detecting the most massive galaxies in this epoch, does not contradict simulations, but rather requires that adjustments be made to high-z modelling of star formation. Overall, the masses from previous detection of $z>10$ galaxies and the new MoM-z14 and GS-z14 detections agree more with the predicted masses of the RP region rather than the Normal region, possibly highlighting the need to revise how we treat star formation in the most massive halos in the early Universe.

In Figure \ref{fig:lf}, we plot the UV luminosity functions of the simulated and measured galaxies. We use the UV luminosities calculated at $z = 15$ using \renaissance{} from \cite{osheaPROBINGULTRAVIOLETLUMINOSITY2015a}, which were derived from the stellar masses, star formation rates (SFRs) and metallicities of galaxies from each region. We refer the reader to \citep{mccaffreyNoTensionJWST2023c} for more information on how SFR values derived from \renaissance{} compare to previous JWST observations. There are concerns that there is an overabundance of luminous galaxies in the early Universe, suggesting high star formation efficiencies at this epoch \citep{SFE-concerns, sfe-concerns-2, sfe-concerns-3}. Studying and comparing the UV luminosities from \renaissance{} with the values derived from observations and new theoretical models will be valuable in understanding where inaccuracies remain in the modelling of star formation at high-z. \\
\indent In Figure \ref{fig:lf} we can see that the measured UV luminosity function from the combination of GS-z14 and MoM-z14 lie at lower values than the simulations, with considerably larger UV magnitudes, due to the large respective star formation rates of $25_{-5}^{+6}$ and $13_{-3.5}^{+3.7}$ \msolaryr for these galaxies. The luminosity function values that \cite{naiduCosmicMiracleRemarkably2025} calculate in their work comes from the assumptions that these are the only $z\sim14$ galaxies to exist in the FoV of the survey.

The UV luminosity functions from \renaissance{} fit well to the luminosity function calculations at $z = 14$ from the adjustments made by \cite{mauerhoferSynergisingSemianalyticalModels2025}, suggesting that \renaissance{}, for the most part, is agreeing with theoretical models that suggest adjustments need to be made in terms of the star formation of the first generation of stars. Where \renaissance{} is most likely breaking down is for the brightest galaxies (as found by JWST) where the rapid assembly and subsequent evolution driving these galaxies is not well captured by our physical models and/or resolution. \renaissance{} still struggles to achieve the brightness of the measured galaxies. This could possibly be explained by the fact that the UV luminosity function derived from GS-z14 and MoM-z14 has a measured value of $\sim 10^{-5}$ cMpc$^{-3}$ mag$^{-1}$. If \renaissance{} were to achieve such a statistic, it would require a volume far greater than it currently has.

\section{Conclusions}
In conclusion, \renaissance{} is underestimating the stellar masses of the most massive high-z galaxies with the majority of them fitting the RP line from Fig.\ref{fig:stellar-mass} but not the Normal region (as may be expected). The misalignment between the Normal region and the observed $z>10$ galaxies indicates that some adjustments need to be made to the star formation/feedback prescriptions and/or numerical resolution needed in future simulations to model these high-z, extreme, galaxies. \\
\indent It is unlikely, given the relatively small field of view of the JADES and MoM surveys that JWST by chance, happened to set its sights on a RP-like region in the early Universe. While the UV luminosities from the observations line up with the derived quantities from \renaissance{}, suggesting that the global star formation rate is accurately modelled in the simulations, the high masses of these individual galaxies themselves may suggest that there needs to be some revisions on how we apply stellar formation prescriptions in select environments \citep[e.g.][]{Calura22_Siege, Brauer25_Aeos}. These select environments may allow, for example, earlier star formation in rapidly accreting galaxies to account for this increase in galaxy size or even greater star formation efficiency for the most massive galaxies at high-z. We also refer to the ongoing issue in cosmological simulations to resolve small areas of bursty, brief star formation \citep{bursty-SF-2, bursty-SF-1}, which may account for the underestimation of star formation in early galaxies. There are also studies that are able to explain the high-z galaxy properties, without the need for an enhanced SFE \citep{dust-attenuation}, through a decreasing dust attenuation model and also simulations that use a top-heavy stellar IMF \citep{top-heavy-IMF-2, top-heavy-IMF-1}. 

We conclude that there is insufficient evidence to imply that our most accepted model of the Universe, $\Lambda$CDM, is in tension with the observations that JWST has been making for three years now. There is still not enough data relating to $z>10$ galaxies to arrive at the conclusion that we need to change the most fundamental foundations of our current framework of galaxy formation and cosmology.

\section*{Acknowledgements}
\noindent JM acknowledges the support from the John \& Pat Hume Doctoral Awards Scholarship (Hume 2021-22).  JHW acknowledges support by NSF grants AST-2108020 and AST-2510197 and NASA grant 80NSSC21K1053.  JR acknowledges support from the Royal Society and Research Ireland under grant number URF\textbackslash R1\textbackslash 191132 and from the Laureate programme under grant number IRCLA/2022/1165. We also thank the referee for supplying us with useful feedback.

\bibliographystyle{aasjournal}
\bibliography{references}

\end{document}